# Now, Later, and Lasting: Ten Priorities for AI Research, Policy, and Practice[†]

Eric Horvitz     Vincent Conitzer     Sheila McIlraith     Peter Stone

April 2024

Advances in artificial intelligence (AI) will transform many aspects of our lives and society, bringing immense opportunities but also posing significant risks and challenges. The next several decades may well be a turning point for humanity, comparable to the industrial revolution. If so, future historians will judge how well we harnessed the benefits of AI for humanity, while protecting against potential harms. We write to share a set of recommendations for moving forward from the perspective of a founder and leaders of the One Hundred Year Study on AI.[1] Launched a decade ago with a dedicated endowment, the project is committed to a perpetual series of studies by multidisciplinary experts to evaluate the immediate, longer-term, and far-reaching effects of AI on people and society,[2] and to make recommendations about AI research, policy, and practice.[3,4] Beyond these recurrent studies and reports, our initiatives have included related efforts aimed at providing a diverse audience with insights about the trajectory of AI, including the creation of the AI Index, an annual benchmarking of AI progress.[5]

As we witness new capabilities emerging from neural models, it is crucial that we engage in efforts to advance our scientific understanding of these models and their behaviors. We must address AI's impact on people and society through technical, social, and sociotechnical lenses, incorporating insights from a diverse range of experts including voices from engineering, social, behavioral, and economic disciplines. By fostering dialogue, collaboration, and action among various stakeholders, we can strategically guide the development and deployment of AI in ways that maximize its potential for contributing to human flourishing.

Despite the growing divide in the field between focusing on short-term versus long-term implications, we think both are of critical importance. As Alan Turing, one of the pioneers of AI, wrote in 1950, "We can only see a short distance ahead, but we can see plenty there that needs to be done."[6] We offer the following ten recommendations for action that collectively address both the short- and long-term potential impacts of AI technologies:

1. **Scientific Foundations.** Invest deeply in research to extend the scientific understanding of the capabilities and limitations of AI systems. Aim to ascertain the principles underlying their

---

behaviors, identify and address deficiencies, and enable robust human guidance and oversight. Explore a diversity of alternative and complementary AI techniques and prioritize the study of the basis for sudden advances in AI capabilities, such as emergent phenomena observed in neural models as they are scaled.

2. **Safety, Reliability, Equity.** Boost investments in R&D on identifying and mitigating concerns with safety, fairness, accuracy, and reliability of models, and encourage greater sharing and transparency from organizations building the largest-scale models, particularly regarding their methods, practices, results, and experiences. Advances in technology and policy can foster trust, accountability, and collaboration among the AI community and society at large.

3. **Regulation.** Devise best practices, audits, and laws and regulations to incentivize reporting on, modifying, and addressing new capabilities and emergent phenomena to ensure that fielded models and systems are safe and responsibly deployed. These efforts can lead to appropriate oversight, feedback, and governance mechanisms aimed at aligning AI systems with human values and interests.

4. **Disinformation.** Address uses of AI by malicious actors for disinformation, manipulation, and impersonation, which pose a risk to an informed citizenry and to the fabric of our democracies. Beyond harnessing promising technical and sociotechnical strategies built on content provenance, watermarking, and fingerprinting, investigate new laws, regulations, and enforcement to combat these threats while protecting civil liberties. The goal is to safeguard the integrity of information and communication against malevolent repurposing of widely available AI methods to deceive and persuade.

5. **Resources.** Bridge the academia-industry gap to ensure greater access and enhanced collaboration for scholars and students so they can participate fully in frontier research and address key issues when they first arise rather than when it is too late. Address significant asymmetries in resources and access, spanning data, computation, and analytical tools and models, between academic and private-sector research scientists. Nurturing university-based research promises to assure the diversity, quality, and relevance of AI research and innovation.

6. **Vibrancy of Academy.** Mitigate the brain drain of faculty and top potential students to industry by creating incentives and support structures to maintain the health and vibrancy of our educational institutions. Cultivating and retaining top computer-science talent at universities will be critical for research advancement and for teaching and mentoring the next generation of AI leaders.

7. **Jobs and Economy.** Rigorously monitor the influences of AI technologies on jobs and the economy, and creatively pursue technologies and policies that promote shared prosperity rather than increased economic disparity. Prioritize research and development on innovations that bring AI and people together in new forms of collaborative work and that support the primacy of human agency and contribution. These efforts can foster social and economic inclusion and empowerment in the face of AI-driven automation and transformation.

8. **Diplomacy.** Stand up and nurture international diplomatic efforts centered on AI, its implications, and uses. Engage with all nations, including authoritarian and autocratic governments, to jointly promote AI safety and protect human rights. Establish and incentivize the adoption of global norms for responsible AI practices, including recommended practices and prohibitions on uses of AI that jeopardize human rights. The aim is to foster global cooperation and coordination on AI issues that transcend national boundaries and interests.

9. **Deep Currents.** Invest deeply in understanding and tracking subtle but potentially profound longer-term influences with psychological, social, and cultural impacts of AI technologies, such as human self-identity, agency, potential overreliance on AI, and psychological dependence on AI companion systems, as well as other emerging issues associated with new forms of intelligent systems in our daily lives. Engage in focused interdisciplinary research efforts to anticipate and address human and societal implications of AI that may not be immediately apparent or measurable.

10. **Catastrophic Outcomes.** Prioritize the scrutiny of potential catastrophic outcomes associated with AI. Adopt a methodical, scientific approach to investigate these risks. Establish forward-looking strategies that incorporate failsafe designs, crucial safeguards, and preventative measures against severe disruptions. Anticipate, simulate, and "red team" threats stemming from unforeseen consequences of well-intentioned and malicious uses of AI. This includes the possibilities of using AI-powered protein design to create fast-spreading pathogens and trajectories leading to inescapable dependencies or irrevocable losses of human control over AI systems. Maintain constant vigilance via modeling and monitoring of potential pitfalls to ensure safety. Formulate effective oversight mechanisms and best practices. Introduce laws and regulations that preclude deployment of powerful AI systems that do not meet standards of due diligence. Proactive efforts on catastrophic risks can help to prevent and mitigate worst-case scenarios of AI.

We recognize that these recommendations are high-level and broad, and that they do not address all the nuances and complexities of the issues. They are not meant to provide definitive answers or solutions, but rather to spark and guide discussions and actions among the AI community and society at large.

We must lean in vigorously on the *now* and the *soon* to address both short- and long-term issues to shape the future of AI for the common good. We can learn from what has gone well and poorly in previous technological revolutions, as well as assess what is unique about this one. We have the opportunity and the responsibility to invent and design our future, as we always have. This is not a passive or predetermined process, but an active and creative one, that depends on our choices, actions, and collaborations. By working together from our different perspectives and disciplines, we can shape the future of AI for the common good, and for the flourishing of generations to come.

**References**

1. One Hundred Year Study on Artificial Intelligence homepage, Stanford University, https://ai100.stanford.edu

2. Horvitz, E. (2014), One-Hundred Year Study on Artificial Intelligence: Reflections and Framing, 2014, https://ai100.stanford.edu/about/reflections-and-framing

3. Stone, P., Brooks, R., Brynjolfsson, E., Calo, R., Etzioni, O., Hager, G., Hirschberg, J., Kalyanakrishnan, S., Kamar, E., Kraus, S. and Leyton-Brown, K. (2016), Artificial intelligence and Life in 2030: The One Hundred Year Study on Artificial Intelligence. *arXiv preprint arXiv:2211.06318*. https://arxiv.org/abs/2211.06318

4. Littman, M.L., Ajunwa, I., Berger, G., Boutilier, C., Currie, M., Doshi-Velez, F., Hadfield, G., Horowitz, M.C., Isbell, C., Kitano, H. and Levy, K. (2021). Gathering Strength, Gathering Storms: The One Hundred Year Study on Artificial Intelligence (AI100) 2021 Study Panel Report. *arXiv preprint arXiv:2210.15767*. https://arxiv.org/abs/2210.15767

5. Maslej, N., Fattorini, L., Perrault, R., Parli, V., Reuel, A., Brynjolfsson, E., Etchemendy, J., Ligett, K., Lyons, T., Manyika, J., Niebles, J. C., Shoham, Y., Wald, R., Clark, J., The AI Index 2024 Annual Report (2024). https://aiindex.stanford.edu/report/

6. A. M. Turing (1950), Computing Machinery and Intelligence, *Mind*, New Series, Vol. 59, No. 236, October 1950, Oxford University Press, pp. 433-460 (p.460).

---

**About the Authors**

*Eric Horvitz* is cofounder of the One Hundred Year Study on AI. He is the Chief Scientific Officer of Microsoft, a member of the President's Council of Advisors on Science and Technology (PCAST), and Chair Emeritus of the Partnership on AI.

*Vince Conitzer* is the chair of the Standing Committee of the One Hundred Year Study on AI. He is a professor of computer science at Carnegie Mellon University and Head of Technical AI Engagement at the Institute for Ethics in AI at the University of Oxford.

*Sheila McIlraith* is the chair-elect of the Standing Committee of the One Hundred Year Study on AI. She is a professor of computer science at the University of Toronto, a Canada CIFAR AI Chair (Vector Institute), and Associate Director, Schwartz Reisman Institute for Technology and Society.

*Peter Stone* is the past-chair of the Standing Committee of the One Hundred Year Study on AI. He is a professor of computer science at the University of Texas at Austin and Executive Director of Sony AI America.